\documentclass[twocolumn,showpacs,amsfonts,amssymb,amsmath,floats,aps]{revtex4-1}
\usepackage{graphicx}
\usepackage{dcolumn}
\usepackage{multirow}
\usepackage{braket}
\usepackage{bbold}
\usepackage{color}
\usepackage{amsmath,amssymb}

\usepackage[normalem]{ulem} 

\usepackage[outdir=./]{epstopdf}
\usepackage{upgreek}
\usepackage{nicefrac}

\def\ket#1{|#1 \rangle}

\def\w{\omega}

\usepackage{xcolor}

\definecolor{mygreen}{RGB}{36,130,25}

\DeclareGraphicsExtensions{.pdf,.eps,.png}

\begin{document}

\title{Fourth-order spin correlation function in the extended central spin model}

\author{Nina Fr\"ohling}
\thanks{These two authors contributed equally.}
\address{Lehrstuhl f\"ur Theoretische Physik II, Technische Universit\"at Dortmund,
Otto-Hahn-Stra{\ss}e 4, 44227 Dortmund, Germany}

\author{Natalie J\"aschke}
\thanks{These two authors contributed equally.}
\address{Lehrstuhl f\"ur Theoretische Physik II, Technische Universit\"at Dortmund,
Otto-Hahn-Stra{\ss}e 4, 44227 Dortmund, Germany}

\author{Frithjof B.\ Anders}
\address{Lehrstuhl f\"ur Theoretische Physik II, Technische Universit\"at Dortmund,
Otto-Hahn-Stra{\ss}e 4, 44227 Dortmund, Germany}

\date{\today}

\begin{abstract}

Spin noise spectroscopy has developed into a very powerful tool to access the electron spin dynamics. While the spin-noise power
spectrum in an ensemble of quantum dots in a magnetic field 
is essentially understood, we argue
that the investigation of the higher order cumulants  promises to provide additional information
not accessible by the conventional  power noise spectrum.
We present a quantum mechanical approach to the correlation function 
of  the spin-noise power
operators at two different frequencies for small spin bath sizes and compare the results with
a simulation obtained from the classical spin dynamics for large number of nuclear spins.
This bispectrum is defined
as a two-dimensional frequency cut in the parameter space of the fourth-order 
spin correlation function. It reveals 
information on the influence of the nuclear-electric  quadrupolar interactions on
the long-time electron spin dynamics dominated by a magnetic field. For large bath sizes
and spin lengths the quantum mechanical spectra converge to those of the classical simulations.
The broadening of the bispectrum across the diagonal in the frequency space is a direct measure
of the quadrupolar interaction strength. A narrowing is found with increasing magnetic field indicating a suppression of the influence of quadrupolar interactions in favor of the nuclear Zeeman effect.

\end{abstract}
\maketitle

\section{Introduction}

Optical spin noise spectroscopy (SNS) \cite{SinitsynPershin} has been
established as a minimally invasive probe to study the electron spin
dynamics and was originally proposed by Aleksandrov and Zapasskii
\cite{aleksandrov81,Zapasskii:13}. Off-resonant Faraday rotation
measurements were used for nearly perturbation-free measurement of the
spin noise in an ensemble of alkali atoms \cite{Crooker_Noise}, as
well as in bulk semi-conductors \cite{Oestreich_noise,
PhysRevB.79.035208, PhysRevLett.115.176601}.  In the absence of an
external magnetic field, SNS was able to reveal the influence of the
electrical-nuclear quadrupolar interactions on an ensembles of
semi-conductor quantum dots (QD) \cite{RevModPhys.79.1217} onto the
long-time decay \cite{FinleyNature,UhrigHackmann2014} of the
second-order spin correlation function $C_2(t)=\langle
S_z(t)S_z\rangle$ as well as on its spin-noise power spectrum
\cite{Crooker2010,PhysRevLett.108.186603, Glasenapp2016,
PhysRevB.89.045317, PhysRevLett.115.207401, PhysRevB.93.035430,
PhysRevLett.109.166605}.

The information extracted from the second order correlation function,
however, is limited to macroscopic linear effects by the fluctuation
dissipation theorem, if only the thermal equilibrium is
considered. For this reason many experimental studies utilized
non-equilibrium conditions, generated by radio frequency
\cite{Glasenapp2014, PhysRevLett.102.087604, PhysRevLett.99.250601} or
through periodic laser pulses \cite{Young_2002_C2, JaeschkeGlazov2018,
PhysRevLett.96.227401, Greilich2006, PhysRevB.96.205419, Varwig2016}.

Applying an external magnetic field to a QD whose strength is
exceeding the Overhauser field \cite{PhysRevB.65.205309,Testelin2009}
generated by the surrounding fluctuating nuclear spins, however,
suppresses the effect of these quadrupolar interactions onto $C_2(t)$
\cite{Glasenapp2016}.  Yet, quadrupolar interactions play an important
role in the understanding of the long-time decay of higher-order spin
response functions \cite{Press2010,Bechtold2016,PhysRevB.96.045441}
especially at large magnetic fields above 1T. Since those fourth-order
spin correlation functions
\cite{Press2010,Bechtold2016,PhysRevB.96.045441} have been
investigated in the time domain, it has been suggested
\cite{1367-2630-15-11-113038,LiSinitsyn2016} to extend the
conventional SNS to higher-order spin noise correlations.  The
third-order spin correlation of the type $C_3(t_1, t_2, t_3)=\langle
S_z(t_1)S_z(t_2)S_z(t_3)\rangle$ requires time-reversal symmetry
breaking for non-zero values, is imaginary in the time domain in
thermal equilibrium and is, therefore, not an observable.
 
In this paper, we focus on the spectral information of the
fourth-order correlation functions \cite{LiSinitsyn2016,
1367-2630-15-11-113038, PhysRevB.96.045441} in the weak measurement
regime.  While it has been established that the real-time fourth order
spin correlation functions contain important information on the
field-dependent long-time scale of the decay
\cite{Press2010,Bechtold2016} that was connected to competition
between the nuclear quadrupolar couplings and the Zeeman energy \cite{
PhysRevB.96.045441}, not much is know on the spectral information of
the fourth order correlation function \cite{LiSinitsyn2016,
1367-2630-15-11-113038}.
 
In a recent publication H\"agele et al.\ proposed an approach to
higher-order cumulants and their spectra in the context of a
continuous quantum noise measurement \cite{PhysRevB.98.205143}. The
authors used a time evolution of the density operator by combining the
von Neumann equation with a Markovian damping through the environment
and a feedback term generated by the continuous measurement of the
property of interest. They applied their method to calculate the
fourth-order correlation spectrum for two coupled spins in a finite
magnetic field.

We present and compare two far simpler approaches based on the linear
response theory that is tailored to the nearly perturbation free
optical SNS \cite{LiSinitsyn2016,
1367-2630-15-11-113038,Crooker2010,PhysRevB.79.035208} and is easily
applicable to a wide range of scenarios. We are interested in the
correlations between the spin-noise power spectrum operator at two
different frequencies. For independent noise variables and purely
Gaussian noise, the correlation function would factorize and the
cumulant \cite{doi:10.1143/JPSJ.17.1100} would vanish.  If the spin
dynamics is coherent, this fourth-order spin correlation function
yields only non-vanishing contributions at the same frequencies.

We focus on the spin dynamics in a semi-conductor quantum dot since it
is considered as promising candidate for a quantum bit
\cite{Bennett2000, Burkard2000QDComputer, DiVincenzo1998}.  We analyze
the influence of the spin length and number of nuclear spins onto the
fourth-order spin noise spectra using the central spin model (CSM)
\cite{Gaudin1976} and its extension to nuclear electric quadrupolar couplings
\cite{PhysRevLett.109.166605,PhysRevLett.115.207401} which is well
suited to describe quantum dot systems
\cite{PhysRevB.65.205309,PhysRevB.70.195340,AlHassanieh2006,Glazov2012}.
We show that the spectra calculated by our quantum mechanical method
approach the results obtained by a semiclassical simulation
\cite{ChenBalents2007,PhysRevB.96.205419} in the limit of larger spins
and bath sizes. In the opposite limit, we are able to reproduce the
higher-order spin spectra in the case of two coupled spins in a finite
magnetic field presented in Ref.~\cite{PhysRevB.98.205143}.

The main obstacle for the realization of a quantum
bit~\cite{Bennett2000, Burkard2000QDComputer, DiVincenzo1998} by a QD
ensemble is the loss of information, as the electron spin decays over
time due to its coupling to a fluctuating environment. While spin
decoherence due to free electron motion is suppressed in a quantum
dot, the high localization causes the hyperfine interaction between
the electron spin and the surrounding nuclear spins to dominate.  A
detailed investigation of interaction processes influencing the spin
dynamics in quantum dots on all time scale is desirable.

The standard SNS was successfully established as a very useful tool to
obtain the basis information on the spin dynamics. However, spectral
information on very weak interactions such as the nuclear quadrupolar
interactions or the influence of the dipole-dipole interactions is
lost rather quickly in a finite magnetic field larger than the
Overhauser field. We propose to investigate the fourth-order spin
noise spectrum since the shape of the cumulant spectra is
significantly altered in the presence of the nuclear quadrupolar
interactions even in larger external magnetic fields. Therefore, the
fourth-order spin noise spectrum reveals additional information on the
long time dynamics that is not accessible with the standard SNS. We
are able to link the change in the spectroscopic data to the magnetic
field dependency of the long-time decay time in higher-order
correlations functions \cite{FinleyNature,
Press2010,PhysRevB.96.045441}.

A short introduction of the model and its semi-classical approximation
in Sec.~\ref{sec:model} is followed by the definition of the second
and fourth-order cumulant of the electron spin in
Sec.~\ref{sec:CorrFct}. Quantum mechanical and classical
implementation of the higher-order correlations are presented in
Sec.~\ref{sec:methods}. The results are discussed in
Sec.~\ref{sec:results}. The influence of the external magnetic field
on the spin cumulant is investigated and the classical simulation is
discussed as a limit to the quantum-mechanical approach. The
quadrupolar interaction is included into the model. Classical and
quantum mechanical treatment will provide an insight into its effect
on the higher order spectrum. A brief summary is given in
Sec.~\ref{sec:Conclusion}.

\section{Model}

\label{sec:model}

While an electron spin in a singly negatively charged QD is well
isolated from decoherence due to fluctuating charge environments, the
strong localisation of the electronic wave function increases the
coupling between the electron spin and its surrounding nuclear
spins. The spin dynamics in the QD is governed by interactions acting
on vastly different time scales ranging from the hyperfine interaction
($\sim 1$\,ns) \cite{PhysRevLett.109.166605, PhysRevLett.115.207401}
to the dipole-dipole interaction ($\sim 100$\,$\upmu$s)
\cite{PhysRevB.65.205309}. The quadrupolar interaction caused by
electric strain fields and the nuclear spin depends highly on sample
growth. We will confine ourselves to the dominating interactions for
the spin dynamics in a quantum dot: The hyperfine interaction, the
Zeeman interaction of the spins with the external magnetic field
$\vec{B}_{\text{ext}}$, and the nuclear-electron quadrupolar
interaction.

\subsection{Central spin model}

\label{sec:CSM}

Both the hyperfine interaction as well as the Zeeman energy
of the spins in an external magnetic field $\vec{B}_{\text{ext}}$ is described by the Hamiltonian of the central spin model \cite{Gaudin1976}: 
\begin{align}\label{eqn:Hamiltonian_CSM}
\begin{split}
\tilde H_{\text{CSM}} = g_{e}\mu_{B}\vec{B}_{\text{ext}}\vec{S} +\mu_{N}\vec{B}_{\text{ext}} \sum_{k=1}^N 
g_{N,k}\vec{I}_k + \sum_{k=1}^N  \tilde{A}_k \vec{I}_k\vec{S}.
\end{split}
\end{align}
 $g_{e}$ denotes the g-factor of the electron, $g_{N,k}$ accounts for the g-factor of the $k$-th nucleus, and $\mu_{N}$ is
 the nuclear magneton. The third term represents the hyperfine interaction between the electronic central spin $\vec{S}$ and the bath comprised $N$ nuclear spins $\vec{I}_k$ conveyed by the coupling constants $\tilde{A}_k$. In negatively charged QDs the hyperfine interaction is isotropic \cite{PhysRevB.89.045317}. The fluctuation frequency
 \begin{align}
 \label{eqn:w-fluc}
 \omega^2_{\mathrm{fluc}}=\frac{4}{3} \langle \vec{I}^2 \rangle\sum_{k=1}^N\tilde{A}_k^2
 \end{align}
 of the Overhauser field
\begin{align}
\vec{B}_N = \sum_{k=1}^N \tilde{A}_k \vec{I}_k
\end{align}
is used to define the intrinsic time scale $T^{\ast}=1/\omega_{\mathrm{fluc}}$ which describes the short-time electron spin decoherence induced by the fluctuation of the
Overhauser field. Typical values of 1-3\,ns are found for $T^{\ast}$ in experiments, depending on their lateral size \cite{PhysRevLett.115.207401,FinleyNature}. The expectation value of the nuclear spin length $\langle \vec{I}^2 \rangle$ takes the value $ I(I+1)$  in the quantum mechanical case
while for the  classical approximation, we obtain the spin length $\langle \vec{I}^2 \rangle = I^2$.

(In,Ga)As/GaAs QDs contain different
isotopes with different spin lengths. While Ga and As isotopes
are characterized by a nuclear spin of $I_{Ga} = I_{As} =
3/2$, In has a spin $I_{In} = 9/2$. Therefore, the influence of different $I$ on higher order correlation functions will be discussed at length in this paper.

The time scale $T^*$ can be utilized to introduce a dimensionless Hamiltonian
\begin{align}
H=T^{\ast}\tilde H_{\text{CSM}}
\end{align}
with the dimensionless hyperfine coupling constants $a_k=T^{\ast}\tilde A_k$
and the dimensionless external magnetic field $\vec{b}_\mathrm{ext} = T^{\ast}g_{e}\mu_{B}\vec{B}_{\text{ext}}$. 
Assuming that all nuclear spins have the same $g$-factor it is convenient to define 
\begin{align}
\zeta = \frac{g_{N}\mu_{N}}{g_{e}\mu_{B}},
\end{align}
the ratio between nuclear and electron Zeeman energy. Then, the Hamiltonian takes the dimensionless form
\begin{align}\label{eqn:Hamiltonian_CSM_dimless}
\begin{split}
H_{\text{CSM}}  =& \vec{S} \vec{b}_\mathrm{ext} + \zeta\vec{b}_\mathrm{ext} \sum_{k=1}^N   \vec{I}_k +\sum_{k=1}^N  a_k \vec{I}\vec{S}.
\end{split}
\end{align}

The hyperfine coupling constants $\tilde A_k$ are proportional to the probability of the electron  at the location of the $k$th nucleus, $\tilde A_k \propto |\psi_e(\vec{R}_k)|^2$. We assume the envelope of the electron wave function in a $d$-dimensional quantum dot with the radius $L_0$ is of the form
\begin{align}
\label{wavefunction}
\psi_e(\vec{r})= CL_0^{-{d}/2}\exp \left(-\frac{|\vec{r}|^m}{2L_0^m}\right),
\end{align}
with $m=1$ describing a hydrogen-like and $m=2$ a Gaussian envelope function. $C$ is a dimensionless normalization constant. With the coupling constant dependent on the probability of an electron
being present at the position of the $k$th nucleus, $\tilde A_k \sim |\psi|^2$, the realization of hyperfine coupling constants thus becomes
\begin{align}
\tilde A_k = A_{\text{max}}\exp \left(-\frac{|\vec{r}|^m}{L_0^m}\right).
\label{eq:Ak_gen}
\end{align}

Due to the growth strain in an semiconductor quantum dot the quadrupolar moment of the nucleus interacts with the strained electronic charge distribution in the QD. In case of axial symmetry regarding the local easy axis $\vec{n}_k$, the quadrupolar interaction is represented by the Hamiltonian 
\cite{PhysRevLett.109.166605,PhysRevLett.99.037401,PhysRevB.85.115313}
\begin{align}
\label{eqn:QPI_hamiltonian}
H_{Q} = \sum_k H_{q}^k = \sum_k q_k (\vec{I}_k \vec{n}_k)^2.
\end{align}
The quadrupolar interaction constants $q_k$ are a measure of the quadrupolar interaction strength at the $k$th nucleus and are quantified by the second derivative of the electron strain field along the easy axis.
The local easy axis $\vec{n}_k^z$ have been reported to be at a mean deviation angle of $\overline{\theta}=23^\circ$ with the growth axis~\cite{PhysRevB.85.115313} for an In$_{0.4}$Ga$_{0.6}$As QD.

\subsection{Semi-classical approximation}
\label{sec:SemiClassical}

In the semiclassical approximation, we replace the quantum mechanical
spin operator with classical vectors and average over all possible
initial spin configurations \cite{ChenBalents2007,PhysRevB.65.205309,
Stanek_Class_2014}.  In numerical simulations, the integral over all
Bloch spheres are replaced by the discrete configuration sample that
introduces some finite statistical error that is well controlled by
the number of configurations.

The basis of the classical simulation is a set of coupled equations of motion for the central electronic spin $\vec{S}$ and the individual nuclear spins $\vec{I}_k$ \cite{PhysRevB.65.205309, Stanek_Class_2014}. Those can be derived as the classical limit from the quantum mechanical Hamiltonian Eq.~\eqref{eqn:Hamiltonian_CSM_dimless} via a path integral formalism \cite{ChenBalents2007, AlHassanieh2006}.
By solving
\begin{align}
\label{eqn:classical_EOM}
\dfrac{\mathrm{d}\vec{S}}{\mathrm{d}t} &=  \left(\vec{b}_\mathrm{ext} + \sum_k a_k \vec{I}_k \right) \times \vec{S} = \vec{b}_{\mathrm{tot}, S} \times \vec{S}, \\
\dfrac{\mathrm{d}\vec{I}_k}{\mathrm{d}t} &=  \left(\zeta\vec{b}_\mathrm{ext} +  a_k \vec{S} \right) \times \vec{I}_k = \vec{b}_{\mathrm{tot}, I_k} \times \vec{I}_k  
\label{eq:cl-Ik}
\end{align}
for different realizations of the initial spin state from the nuclear Gaussian sample space, we can infer the dynamics of the spin expectation values by averaging over the dynamics in each configuration. The mean values of the spin dynamics are interpreted as the time average over consecutive measurements on a single quantum dot.

The dynamics of the electron spin is governed by the external magnetic
field as well the hyperfine interaction with the nuclear spins. Those
two effects can be merged to one time dependent effective field
$\vec{b}_{\mathrm{tot}, S}$ around which the electron spin
precesses. The same holds for the differential equations of the
nuclear spins which are influenced by the nuclear Zeeman term
$\zeta\vec{b}_\mathrm{ext}$ and the Knight field $a_k\vec{S}$.

The classical formalism can also be extended to include the
quadrupolar effects on the nuclear spins \cite{PhysRevLett.99.037401}.
Using the Heisenberg equation with $H_Q$ stated in
Eq.~\eqref{eqn:QPI_hamiltonian} and assuming commuting classical
variables, the effective field $\vec{b}_{\mathrm{tot}, I_k}$,
\begin{align}
\label{eqn:btotIk} 
\vec{b}_{\mathrm{tot}, I_k} =
\zeta\vec{b}_\mathrm{ext} + a_k \vec{S} + 2 q_k
(\vec{n}_k\vec{I}_k)\vec{n}_k,
\end{align} 
can be extended to also comprise the influence of the
quadrupolar interaction $2 q_k (\vec{n}_k\vec{I}_k) \vec{n}_k$.

The quadrupolar interaction induces an additional precession around
the axis $\vec{n}_k$ for each nuclear spin but with a variable
precession frequency.  The angular frequency is given by the scalar
projection of $\vec{I}_k$ onto $\vec{n}_k$ weighted by $q_k$.  Without
hyperfine coupling this leads to a precession around a constant
$\vec{n}_k$ in which the nuclear Zeeman term acts as a perturbation
for small external magnetic fields.

\section{Correlation functions and noise}
\label{sec:CorrFct}

Kubo \cite{doi:10.1143/JPSJ.17.1100}  pointed out that cumulants play an role in the probability 
theory which is important in quantum mechanical systems as well as in the thermodynamics. The observation
that the  moment generating functional 
\begin{eqnarray}
\left\langle e^{\xi X} \right\rangle &=& \exp\left(\sum_{n=1}^\infty \frac{\xi^n}{n!} \kappa_n\right) = \exp\left(\left\langle e^{\xi X} -1\right\rangle_c \right)
\end{eqnarray}
with the parameter $\xi$
is linked to the exponentiated series of the $n$-the order cumulant $\kappa_n$
of the random variable $X$
had a profound impact for the diagrammatic perturbation theory as well as the analysis of the noise \cite{Mendel}. The subscript $c$
refers to the cumulant average.

This concept can be extended to several random variables which will be replaced by operators in quantum mechanical calculations.
The second order cumulant of the two variables $X_1$ and $X_2$ is defined as
\begin{eqnarray}
\langle X_1 X_2\rangle_c &=& \langle X_1 X_2\rangle - \langle X_1\rangle \langle X_2\rangle
\end{eqnarray}
which is identical to  $\langle X_1 X_2\rangle $ if the mean average $\langle X_1\rangle$  vanishes.
The same principle can be applied to higher orders  \cite{doi:10.1143/JPSJ.17.1100} and is the basis for the spin-noise analysis
presented in this paper. 

Here, we are using the Heisenberg operators $S_z(t)$ as variables to define spin-spin correlation
functions.  In order to access the frequency information for the  spin correlation functions of arbitrary
order, we introduce the  Fourier transformation
\begin{align}
\label{eq:fourier-variable}
a(\omega) = \frac{1}{\sqrt{T_m}}  \int_{-T_m/2}^{T_m/2} \text{d}t\,e^{-\text{i}\omega t}S_z(t),
\end{align}
with the measurement time $T_m$ and the measurement starting at $t_0=-T_m/2$. 

If the measuring time $T_m$ is large compared to the characteristic time scale
of spin decay,   we can apply the limit $T_m\to \infty$ to simplify the mathematical expressions.
Note, however, that one has to be careful when applying this limit to avoid unexpected divergence in expressions.
We  point out below when we have to resort to the original finite measurement time $T_m<\infty$
to remove any ambiguities.

\subsection{Second-order correlation function }
\label{sec:C2}

The spin-noise experiments in semiconductor QD are generally performed at $T=4-6K$, so that
the thermal energy is large compared to the energy scale generated
by the Overhauser field. 
Furthermore,  we can
neglect the equilibrium spin polarization $\langle S_z\rangle$ 
for a sufficiently low external magnetic field
so that the second-order spin spin auto correlation function
is identical to its cumulant.
This second-order auto correlation function 
\begin{align}
\tilde C_2(t_1, t_2) = \langle S_z(t_1)S_z(t_2)\rangle
\end{align}
describes the correlation between the $z$-component of the spin at the start of the measurement $t_1$ and at a time $t_2$. 

Since experiments on spin noise in quantum dots are usually performed
in the linear response regime, we assume that the system is in
equilibrium and the Hamiltonian commutes with the density operator.
This implies that the system is translationally invariant in time, the
correlation function only depends on the relative time $\tau=t_1-t_2$,
and therefore can be expressed as $C_2(\tau)=\langle S_z(\tau)S_z(0)
\rangle$.  This holds for all higher order auto correlation functions:
for systems that are translational invariant in time one time variable
is usually eliminated \cite{Mendel} such that the $k$-th order
correlation function only depends on $k-1$ time variables.

The  Wiener-Chintchin theorem \cite{Khintchine1934, SinitsynPershin}  relates the steady-state spin auto correlation function to
the noise power spectrum. It requires that the measuring time $T_m$ is much longer that the characteristic time scale of the spin
decay $T^*$ ($T_m\gg T^*$). Substituting the Fourier transformation \eqref{eq:fourier-variable} and using the  translational invariance in time,
we obtain the second-order spin correlation function in
the frequency domain:
\begin{eqnarray}\label{eqn:general_C2_def}
\tilde C_2(\omega_1, \omega_2)& =&  \lim_{T_m\rightarrow \infty} \langle a(\omega_1)a(\omega_2)\rangle
\nonumber \\
&=& \lim_{T_m\rightarrow \infty} \frac{1}{T_m} \int_{-\frac{T_m}{2}}^{\frac{T_m}{2}} dt_1 \,e^{-\text{i}\omega_1 t_1}
\int_{-\frac{T_m}{2}}^{\frac{T_m}{2}} dt_2 \,e^{-\text{i}\omega_2 t_2}
\nonumber \\
&&\langle S_z(t_1) S_z(t_2)\rangle
= \delta_{\omega_1,-\omega_2} C_2(\omega)
\label{eq:17}
\end{eqnarray}
with
\begin{align}\label{eq:18}
C_2(\omega) =  \int_{-\infty}^{\infty} d\tau \langle S_z(\tau ) S_z(0)\rangle e^{i\omega \tau}.
\end{align}
$C_2(\omega)$ denotes the spin-noise spectrum and satisfies the sum rule
\begin{align}
\int_{-\infty}^\infty d\omega C_2(\omega) = \frac{\pi}{2}.
\label{eq:sumruleC2}
\end{align}
Note that the inclusion of the prefactor $1/\sqrt{T_m}$ into the definition of the Fourier transformation ensures the convergence
of $\tilde C_2(\omega_1, \omega_2)$. It also leads to the Kronecker delta in the last line of Eq.\ \eqref{eq:17}
\footnote{Strictly speaking, the frequencies $\w_i$ are multiples of $2\pi/T_m$ in a Fourier transformation on
a finite interval $[-T_m/2,T_M/2]$ but are asymptotically dense for $T_m\to \infty$, so that we treat them as
continuous variables with the proper mathematical limit implied.
}.

\subsection{Fourth-order correlation function}
\label{sec:S4}

While the second-order spin correlation has been extensively studied
both in the frequency and the time domain \cite{Crooker_Noise,
FinleyNature, Glasenapp2016}, the properties of fourth-order
correlation functions remain relatively unexplored
\cite{LiSinitsyn2016}.

An $n$-th order cumulant is given by the $n$-th order auto-correlation
function from which all combinations of lower order correlation
functions are subtracted -- see Ref.\ \cite{doi:10.1143/JPSJ.17.1100}
for more details.  The basic idea is to separate the true higher order
correlations from a trivial factorisation. If a system would be fully
characterized by Gaussian noise, all higher order cumulants would
vanish \cite{Mendel}.

One can show that the third-order spin correlation function is imaginary in the time domain and not accessible.
In this paper, we therefore focus on the fourth-order spin correlation function.
Its cumulant provides additional information on the dynamics of the system not yet included  in $C_2$.
The fourth-order cumulant of $a(\omega)$ is defined as
\begin{align}
\begin{split}
\tilde S_4(\omega_1, \omega_2, \omega_3, \omega_4) = & \tilde C_4(\omega_1, \omega_2, \omega_3, \omega_4) \\
&-\tilde C_2(\omega_1, \omega_2)\tilde C_2(\omega_3, \omega_4)\\
&-\tilde C_2(\omega_1, \omega_3)\tilde C_2(\omega_2, \omega_4)\\
&-\tilde C_2(\omega_1, \omega_4)\tilde C_2(\omega_2, \omega_3),
\end{split}
\label{eq:S4allg}
\end{align}
where we neglected the spin polarisation in a finite magnetic field, which is
justified in the high temperature limit.
The translational invariance in time in combination with the limit $T_m\gg T^*$
yields the constraint 
\begin{eqnarray}
\tilde C_4(\omega_1, \omega_2, \omega_3, \omega_4)
&=& \delta_{\omega_1+ \omega_2+ \omega_3+ \omega_4,0}  
\\
&&\times 
C_4(\omega_1, \omega_2, \omega_3, -(\omega_1+ \omega_2+ \omega_3)).
\nonumber
\end{eqnarray}

We are interested in a special case of the fourth order cumulant 
$S_4(\omega_1, \omega_2) =\tilde S_4(\omega_1, -\omega_1, \omega_2, -\omega_2)$.
Since $a(-\w) = a^*(\w)$, it correlates two spin-noise power spectrum component $|a(\w)|^2$ at different
frequencies with each other. Using Eq.\ \eqref{eq:S4allg}, this bispectrum fulfils the relation

\begin{align}
\begin{split}
S_4(\omega_1, \omega_2) =&\tilde S_4(\omega_1, -\omega_1, \omega_2, -\omega_2) \\
=&C_4(\omega_1, \omega_2)-C_2(\omega_1)C_2(\omega_2)\\
&\times (1+\delta_{\omega_1, \omega_2}+\delta_{\omega_1, -\omega_2}).
\end{split}
\label{eq:S4_allg}
\end{align}
with $C_4(\omega_1, \omega_2) =\tilde C_4(\omega_1, -\omega_1, \omega_2, -\omega_2)$. 
In the limit $T_m\rightarrow \infty$, the last two terms in Eq.\ \eqref{eq:S4allg} are zero for $S_4(\omega_1, \omega_2)$ unless $\omega_1=\pm \omega_2$. 

If the two frequency components are uncorrelated, the fourth order
cumulant would vanish.  If the cumulant features anti-correlation,
i.\,e. $S_4(\omega_1, \omega_2)<0$, the observation of a spin
component with the frequency $\omega_1$ decreases the likelihood of
simultaneously observing a spin precession with the frequency
$\omega_2$.

In the long measurement limit, the Fourier transform of $C_4(\omega_1, \omega_2)$ becomes
\begin{align}
C_4(t_1, t_2) = \frac{1}{T_m} \int^{T_m/2}_{-T_m/2} d\tau \langle S_z(t_1+\tau)S_z(\tau)S_z(t_2)S_z\rangle .
\end{align}
This integrand describes the correlation of two $C_2(t_{1/2})$
measurements -- one started at $t=0$, the other started at $\tau$. It
is then averaged over the time delay between both measurements. This
could be implemented in an experimental set-up. It is similar, but not
identical, to the fourth-order correlator $\langle
S_z(t_1)S_z(t_1+t_2)S_z(t_1)S_z\rangle$
\cite{Bechtold2016,PhysRevB.96.045441}. We can therefore expect some
comparable behaviour, such as the sensitivity to quadrupolar
interaction even at high magnetic fields.

It is straight forward to proof the sum rule
\begin{align}
\int_{-\infty}^\infty d\omega_1 \int_{-\infty}^\infty d\omega_2 C_4(\omega_1, \omega_2) = \frac{\pi^2}{4}
\label{eq:sumruleC4}
\end{align}
from the definition of $C_4(\omega_1,\omega_2)$.  Since the
contributions to $S_4(\omega_1, \omega_2)$ containing
$\delta_{\omega_1, \pm \omega_2}$ have the measure zero, the integral
of $S_4(\omega_1, \omega_2)$ over the $\omega_1-\omega_2$-plane
vanishes. This follows from the combination of Eqs.\
\eqref{eq:sumruleC2} and \eqref{eq:sumruleC4}.  Consequently, a
non-vanishing bispectrum must contain as much spectral weight in the
anti-correlations as in the correlations independently of the details
of the Hamiltonian.  Since the term $C_2(\omega_1)C_2(\omega_2)$ in
Eq.\ \eqref{eq:S4_allg} is well understood, the distribution of
correlated and anti-correlated frequencies under the influence of a
transversal magnetic field as well as quadrupolar interaction shall be
the focus of this paper.

\section{Methods}
\label{sec:methods}

In this section we discuss both the quantum mechanical as well as the classical method 
employed for computing second and fourth order correlation functions. 

\subsection{Quantum mechanical approach}

Using an exact diagonalization of the Hamiltonian as a quantum
mechanical approach to the higher order spin correlations suffers from
the exponential growth of the Hilbert space $\mathcal{D} =
\mathrm{dim}(H) = 2(2I+1)^N$ with $N$, the number of nuclear
spins. One can either utilize an approximate treatment of the dynamics
or solve the problem exactly by fully diagonalizing the total
Hamiltonian $H=H_{\text{CSM}}+H_Q$.  With this method the number of
nuclear spins $N$ is limited to a small bath size.

Diagonalizing the Hamiltonian $H_{\text{CSM}}+H_Q$ produces a finite set of discrete eigenvalues and eigenvectors $H\ket{n}=E_n\ket{n}$. 
We use this eigenbase for calculating the spin-spin correlation function 
$C_2(\omega)$ in frequency space from Eq.~\eqref{eq:18}
\begin{align}
\begin{split}
C_2(\omega) =  \frac{2\pi}{D}\sum_{nm} \delta(\omega-(E_n-E_m))|S_{nm}|^2,
\end{split}
\label{eq:C2_full_leh}
\end{align}
 defining the spin operator matrix element
$S_{nm}= \langle{n|S_z|m}\rangle $. 
The spin-noise spectrum $C_2(\omega)$ is positive semidefinite: the matrix elements $|S_{nm}|^2$ contribute
if the excitation energy $E_n-E_m$ coincides with the external probe frequency $\w$.

The fourth-order spin correlation $C_4(\omega_1,\omega_2)$ can be expressed as
\begin{align}
\begin{split}
C_4(\omega_1, \omega_2) = &\frac{4\pi^2}{D}  \sum_{nml}\sum_{k\in U_n}  \delta(\omega_1-(E_n-E_m))\\
&\times \delta(\omega_2-(E_k-E_l))S_{nm}S_{mk}S_{kl}S_{ln}.
\end{split}
\label{eq:C4_full_leh}
\end{align} 
$U_n$ is the subspace of all eigenstates with the same eigenenergy $E_n$.
If the Hamiltonian 
contains solely non-degenerate eigenstates,
the sum over $k$ reduces to a single term $k=n$:
\begin{align}
\begin{split}
C_4(\omega_1, \omega_2) =& \frac{4\pi^2}{D}\sum_{nml} \delta(\omega_1-(E_n-E_m))\\
&\times \delta(\omega_2-(E_n-E_l))|S_{nm}|^2|S_{nl}|^2.
\end{split}
\label{eq:C4_nondegen}
\end{align}
While $C_2$ offers only information on the spin dynamics depending on
one frequency, $C_4$ reveals the interplay between two frequencies,
$\omega_1=E_n-E_m$ and $\omega_2=E_n-E_l$ weighed with the spin matrix
element $|S_{nm}|^2$ and $|S_{nl}|^2$ respectively.  Note that the
delta-functions in the Lehmann representations \eqref{eq:C2_full_leh}
and \eqref{eq:C4_full_leh} imply the limit $T_m\to \infty$. For a
finite measuring time $T_m<\infty$, the delta-functions are broadened
by a width $\propto 1/T_m$.

Combining
Eqs.\ \eqref{eq:C2_full_leh} and \eqref{eq:C4_full_leh}, the bispectrum $ S_4(\omega_1, \omega_2)$ can be expressed as
\begin{align}
\begin{split}
 S_4(\omega_1, \omega_2) =&\frac{4\pi^2}{D}  \left \{ \sum_{nml}\sum_{k\in U_n} \left [ \delta(\omega_1-(E_n-E_m))\right. \right.\\
&\times \left. \delta(\omega_2-(E_k-E_l))S_{nm}S_{mk}S_{kl}S_{ln}\right] \\
&- (1+\delta_{\omega_1, \omega_2}+\delta_{\omega_1, -\omega_2})\\
&\times \left[ \sum_{nm} \delta(\omega_1-(E_n-E_m))|S_{nm}|^2 \right.\\
&\left.\left.  \times \sum_{kl} \delta(\omega_2-(E_k-E_l))|S_{kl}|^2 \right] \right \}.
\end{split}
\label{eq:QM_S4}
\end{align}

\subsection{Classical treatment}
\label{sec:classical-method}

In the quantum mechanical treatment, we used the definition of the
operator $a(\omega)$ and performed the ensemble average by evaluating
the trace over the Hilbert space.

For the classical simulation we proceed in the same manner. There, the
trace is replaced by a configuration average over all initial
conditions \cite{ChenBalents2007,PhysRevB.96.205419}.  The integral
over the Bloch sphere of each spin is approximated by a finite number
of randomly generated spin configurations. We track the time evolution
$S_z(t)$ determined by Eq.~\eqref{eqn:classical_EOM} in each
configuration.

For the case of $\omega_1 = -\omega_2$ and $\omega_4=-\omega_3$, the 
correlation function $\tilde C_4(\omega_1, \omega_2, \omega_3, 
\omega_4)$ can be written as
\begin{align}
C_4(&\omega_1,\omega_2)= \dfrac{1}{N_{C}}\! \sum_{i\in\mathrm{config}} \mathcal{F}C_2^i(\omega_1) \mathcal{F}C_2^i(\omega_2)
\end{align}
In each classical configuration $i$, the Fourier transformation of the 
electron spin correlation $C_2^i(t) = S_z^i(0) S_z^i(t)$ provides building 
blocks for the correlation between the frequencies $\omega_1$ and 
$\omega_2$.

The correlation function  $C_2(\omega_1)$ \cite{PhysRevB.65.205309,Glazov2012}  
that is subtracted from the fourth-order correlator in the cumulant $S_4$, cf. Eq.~\eqref{eq:QM_S4}, is
calculated using
\begin{align}
C_2(\omega_1) = \dfrac{1}{N_{C}} \sum_{i\in\mathrm{config}} 
\mathcal{F}C_2^i(\omega_1).
\end{align}
While $C_2(\omega_1)$ contains $1/3$ of its total spectral weight at low frequencies $\w\ll \w_{\rm fluc}$ 
at $b_{\rm ext}=0$  \cite{PhysRevB.65.205309,PhysRevB.89.045317,PhysRevLett.115.207401},
it becomes Gaussian for $b_{\rm ext} \gg T^*\w_{\rm fluc}$.
The bispectrum $S_4(\omega_1, \omega_2)$ is then computed via Eq.\ \eqref{eq:S4_allg}, analogous to the quantum mechanical method.

\section{Results}
\label{sec:results}

\subsection{Choice of parameters}

For the simulation, physical realities need to be translated into
parameters for the model to best reflect the actual system. While the
number of nuclei in a quantum dot is of the order of $10^4-10^6$,
simulating them all is computationally non-viable. In modelling the
hyperfine interaction between electron and nuclei, we, therefore,
neglect all nuclei whose distance from the electron exceeds a cut-off
radius $R_0$. Following from Eq.\ \eqref{eq:Ak_gen}, the resulting
distribution of hyperfine coupling constants in a QD of the radius
$L_0$ is realized by
\begin{align}
A_k = A_{\text{max}}\exp(-r_0^m \beta^{m/d})
\label{eq:Ak_hack}
\end{align}
with $r_0=R_0/L_0$ and $\beta$ randomly selected from a uniform
distribution, $\beta \sim \mathcal{U}(0,1)$.  Then the set $\{ A_k\}$
is properly normalized such that they always yield the same $\w_{\rm
fluc}$ defined in Eq.\ \eqref{eqn:w-fluc}.  The distribution of Eq.\
\eqref{eq:Ak_hack} was already applied in the
Refs.~\cite{PhysRevB.89.045317, PhysRevLett.115.207401,
PhysRevB.93.035430, Glasenapp2016}.

To generate an adequate representation of the $a_k$ distribution, it
is necessary to adjust the cut-off radius depending on the bath size
to prevent the dynamics being dominated by only a few strongly coupled
nuclear spins.  For small baths ($N<15$) we choose the relative
cut-off radius $r_0=0.8$, while a larger cut-off $r_0=1.5$ is utilized
for large baths. Here, a $(d=3)$-dimensional quantum dot with a
Gaussian electron wave envelope, $m=2$, is studied. To gauge the
influence of quadrupolar couplings on the decay without having to
account for the decay due to the hyperfine coupling distribution,
homogeneous couplings ($A_k=\mathrm{const}$, $R_0=0$) are used as
well.

We average over the Zeeman energies of the isotopes making up an InGaAs QD to estimate the ratio between nuclear and electron Zeeman energy, $\zeta$.  This results in $\zeta=1/800$; the nuclear Zeeman splitting is about three orders of magnitude smaller than the electron Zeeman splitting and perturbative for dimensionless magnetic fields $\vec{b}_\mathrm{ext}$ smaller than $\mathcal{O}(10^2)$.

To quantify the relative quadrupolar interaction, we introduce the dimensionless ratio \cite{Glasenapp2016}
\begin{align}
\label{eqn:QPI_strength}
Q_r = \frac{\sum_k q_k}{\sum_k A_k}
\end{align}
which relates the total quadrupolar interaction strength to the hyperfine coupling strength. 

First, a set $\{\tilde q_k\}$ is  obtained from a uniform distribution $\tilde q_k \in [0.5, 1]$. With a given $Q_r$, the quadrupolar interaction constants $q_k$ are determined via 
\begin{align}
q_k = Q_r\tilde{q}_k\frac{\sum_k A_k}{\sum_k \tilde{q}_k}
\end{align}
to satisfy the relation in Eq.\ \eqref{eqn:QPI_strength}. 
The local easy axes $\vec{n}_k$  \cite{PhysRevB.85.115313}
have been reproduced by generating isotropically distributed vectors and discarding any vector at an angle with the growth axis larger than $\theta_{\text{max}}=34^\circ$, so that the mean angle becomes $\overline \theta=23^\circ$.
The $z$-axis is aligned to the growth axis of the QD, while the external magnetic field is applied transversally, $\vec{b}_{\text{ext}}=b_x\vec{e}_x$  unless otherwise stated.

The delta-distributions in Eq.\ \eqref{eq:QM_S4} are represented by Lorentzians
\begin{align}
\label{eq:lorentzian}
\Gamma(\omega, \Delta E) = \frac{1}{\pi}\frac{\gamma}{(\omega-\Delta E)^2+\gamma^2}
\end{align} 
with a broadening factor $T^{\ast}\gamma=0.01$. This broadening corresponds to a measuring time $T_m=100T^{\ast}$. 
Although, the choice of this rather arbitrary broadening factor influences the magnitude of $S_4(\omega_1, \omega_2)$,
the total spectral weight remains invariant of the broadening.

In an hypothetical quantum mechanical simulation with $10^5$ nuclear
spins, the excitation spectrum entering Eqs.\ \eqref{eq:C2_full_leh}
and \eqref{eq:C4_full_leh} will be dense due to the almost continuous
distribution of the hyperfine couplings $A_k$ in such a large spin
ensemble. In a very small representation of the nuclear spin bath, the
excitation energies become visibly discrete, and the number of
different frequencies are further reduced by the degeneracies in the
absence of an external magnetic field. In order to compensate for this
effect, we generate $N_a$ different sets of $\{ A_k\}$, perform
independent exact diagonalizations leading to a variation of the
excitation spectrum \cite{PhysRevB.89.045317} and average over the
individual spectral functions.  In the limit $N_a\to \infty$ the
excitation spectrum should approach a continuum, at a finite $N_a$,
the Lorentzians \eqref{eq:lorentzian} start to overlap resulting in
smoothed spectra.  We set $N_a=32$ providing a reasonable compromise
between the computational effort and the smoothness of the spectra.

To obtain the equivalent between the quantum mechanical expectation
value and the classical simulation, the averaging over $N_C$ classical
initial spin configurations is necessary. $N_{C} = 10^5$ is considered
a sufficiently large number of configurations to adequately represent
the entirety of the sample space of the spins.  Each configuration
comprises $N=100$ randomly generated nuclear spins and a central spin
that is fully aligned in $z$ direction at $t=0$.

In the simulation all classical spin vectors are of length unity
\cite{PhysRevB.96.205419}. This necessitates the adjustment of the
hyperfine coupling constants $a_k' = S a_k$ and of the Overhauser
field $\vec{b}_\mathrm{N}' = I/S \vec{b}_\mathrm{N}$. It also
translates to the quadrupolar interaction $q_k' = I q_k$.  The
classical spin always represents an effective spin vector
length of $S=I=1/2$ for simplicity.

\subsection{Spin-noise power spectrum $C_2(\w)$}

\label{sec:C2_noise_benchmark}

\begin{figure}[t]

\includegraphics[width=0.39\textwidth, clip]{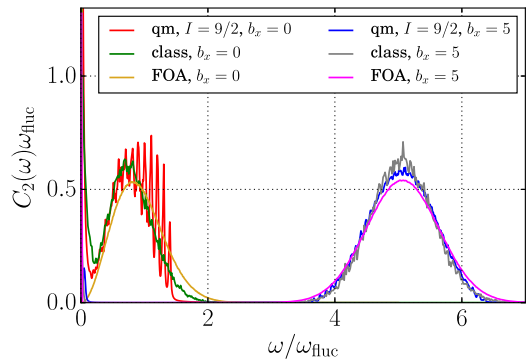}
\caption{Comparison between the spin noise $C_2(\omega)$ for different approaches: solid lines represent the quantum mechanical simulation with three $I=9/2$ nuclear spins, the dashed lines are a classical simulation and the dotted lines are calculated from a Fourier transformed FOA.}

\label{fig:C2_comp}
\end{figure}

To set the stage for higher order spin correlation functions, we
revisit the second order spin noise $C_2(\omega)$ first. A basic
understanding of spin noise was achieved when using the Fourier
transform of the frozen Overhauser field approximation (FOA)
\cite{PhysRevB.65.205309}. The spin noise spectrum was extracted
analytically for $b_x=0$ and numerically calculated for arbitrary
magnetic fields. It was amply discussed in Refs.~\cite{Glasenapp2016,
PhysRevB.89.045317, PhysRevLett.115.207401, UhrigHackmann2014}. In
Fig.~\ref{fig:C2_comp} we provide a comparison of our two methods with
this analytic approximation.  The quantum mechanical and the classical
simulations show good agreement with the solution of the FOA for
$b_x=5$. The deviations of the quantum mechanical result are related
to the small number of simulated bath spins.  At $b_x=0$, the full
spin rotational invariance introduces degeneracies in the
eigenenergies leading to a reduction of the excitation
spectrum. Therefore, the $N_a=32$ different sets of hyperfine coupling
constants are insufficient, and the distinct nuclear frequency peaks
are visible in the spectrum.  This is substantially different at
finite $b_x=5$ where these degeneracies are lifted by the Zeeman
splitting, leading to an almost smooth spectrum.  The classical
simulation traces the Gaussian envelope of the quantum mechanical
spectrum and also differs from the FOA at $b_x=0$.  This is due to the
nuclear spin dynamics included in Eq.~\eqref{eq:cl-Ik} that causes an
additional long-time decay in the time domain not included in the
FOA. Therefore, spectral weight shifts from the delta-peak at
$\omega=0$ to the Gaussian as the non-decaying fraction of $\langle
S_z(t)S_z(0) \rangle$ decreases.

\begin{figure}[tp]
\begin{center}
\includegraphics[width=0.49\textwidth,clip]{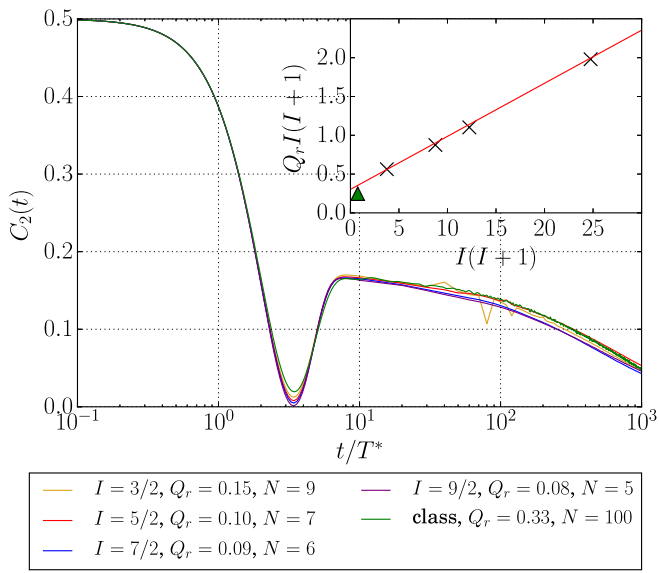}
\caption{The second order spin correlation in the absence of an magnetic field computed by a Lanczos algorithm, with different spin lengths $I$, bath sizes $N$ and interaction strengths $Q_r$, chosen for similar long time decay. The inset plot shows the dependence of $Q_rI(I+1)$ on the squared spin length $I(I+1)$. The hyperfine couplings are homogeneous.}
\label{fig:C2_Idiff}
\end{center}
\end{figure}

When adding the quadrupolar coupling to the central spin model, it is
important to understand its influence on the long-time decay of
$C_2(t)$ as a function of the bath spin length as well as number of
nuclear spins in the simulation. The relative strength $Q_r$ defined
in Eq.\ \eqref{eqn:QPI_strength} has been originally introduced in
Ref.\ \cite{PhysRevLett.115.207401} to minimize this dependency.
Since there is clear experimental evidence
\cite{PhysRevLett.115.207401,Bechtold2016} that $H_Q$ induces a second
long-time decay of $C_2(t)$ which occurs on time scales of $200-600$ns
depending on the growth conditions of the quantum dot ensemble, we aim
for adjusting the value of $Q_r$ for each simulation such that
$C_2(t)$ remains invariant under the change of the bath size or the
spin length in order to maintain a close connection between our
simulations and the experiments.  By establishing this gauge we are
able to compare the differences in the fourth order spectra with
different bath spin lengths $I$ as well as to link the quantum and the
classical simulation.

Figure \ref{fig:C2_Idiff} depicts the second-order correlation
function $C_2(t)$ for different $I$ but similar Hilbert space
dimensions $\mathcal{D}$, with a different but properly adjusted
$Q_r$.  
To make sure that only the
quadrupolar interaction influences the long-time dephasing for $t\gg
T^*$, homogeneous coupling constants $A_k=\mathrm{const}.$ were
chosen. Without quadrupolar interaction the dynamics is equivalent to
the FOA for $b_x=0$.

For $I=3/2$, the quadrupolar coupling strength is set to $Q_r=0.15$, since this value has been successfully used to model experimental data \cite{PhysRevLett.115.207401, PhysRevB.96.045441}. $Q_r$ were chosen for $I=5/2,\,7/2$ and $9/2$ (marked by 'x' in the inset of Fig.\ \ref{fig:C2_Idiff}) so that all 
correlation functions exhibit a similar long-time decay. Interestingly, the $Q_r$ 
that achieve this agreement of $C_2(t)$ for these different combinations of $I$ and $N$ obey the relation
\begin{align}
Q_rI(I+1) = aI(I+1)+b,
\end{align}
with $a=0.068\pm0.002$ and $b=0.30\pm0.03$ obtained via linear regression. The classical computations of $C_2(t)$ that have been made for an 
effective spin vector length of $I=1/2$ follow this relation roughly (marked by a triangle in the inset plot).

\subsection{Fourth-order spin noise in the CSM }

\begin{figure}[tp]
\centering
\includegraphics[width=0.5\textwidth]{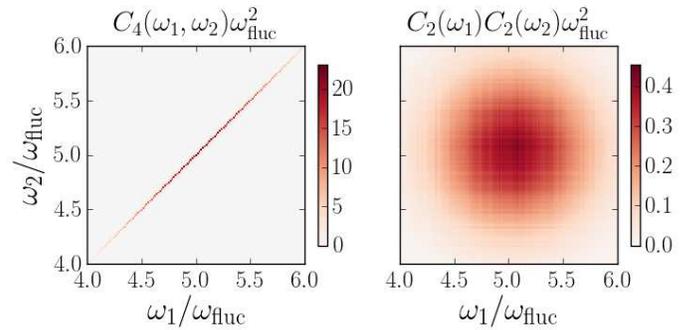} 
\caption{$C_4(\omega_1,\omega_2)$ and $C_2(\omega_1)C_2(\omega_2)$ as the result of a classical simulation. The cumulate $S_4$ spectrum is shown in Fig.~\ref{fig:S4_Ndiff}. The external magnetic field is $b_x = 5$. }
\label{fig:C4_C2_bext5}
\end{figure}

A comparison of the quantum mechanical and classical simulation
results for the fundamental features of the fourth order cumulant
$S_4$ is the topic of this section. We discuss how the shape of $S_4$
is determined by its components $C_4$ and $C_2$ as well as the
dependence of the spectrum on $\vec{b}_\mathrm{ext}$. The classical
simulation is presented as a limiting case to the quantum mechanical
calculation.  To set the stage we limit ourselves for now to the CSM
which excludes the quadrupolar interaction.

\subsubsection{Components of $S_4$ depending on external magnetic field strength}

Each $S_4$ spectrum consists of two parts: $C_4(\omega_1,\omega_2)$
and the product $C_2(\omega_1)C_2(\omega_2)$, cf.\
Eq.~\eqref{eq:QM_S4}.  The results of the classical simulation for
$b_x = 5$ are depicted in Fig.~\ref{fig:C4_C2_bext5}.  Since both
terms only contain quadratic expressions, their individual
contributions are positive.

$C_2(\omega)$ is to good approximation a Gaussian with the mean given
by $\sqrt{b_x^2+1/2}$, cf.~\cite{PhysRevB.89.045317}, and its variance
$\sigma^2$ is determined by the Fourier transform of the envelope of
the central spin dynamics in the time domain for large magnetic fields
$(\omega_\mathrm{fluc}/2)^2)$ \cite{PhysRevB.65.205309}. Since
$\omega_1$ and $\omega_2$ are independent variables, the covariance is
the identity matrix in the multivariate Gaussian given by
$C_2(\omega_1)C_2(\omega_2)$ as shown in the right panel of Fig.\
\ref{fig:C4_C2_bext5}.

$C_4(\omega_1,\omega_2)$ is plotted in the left panel of Fig.\
\ref{fig:C4_C2_bext5}. It only contributes on the diagonal
$\w_1=\w_2$.  This fact is intuitively accessible in the classical
approach. In each configuration the hyperfine interaction changes the
initial frequency given by the generated Overhauser field only
marginally. Therefore, the Fourier transform of $C_2^i(t)$ can be
described by a narrow peak and the product of two distributions can
only be non-zero at the overlap. For better visibility the delta-peaks
are broadened to a Lorentzian with a width of $\gamma T^{\ast}=0.01$.
In the direction of the diagonal, the spectrum follows a Gaussian
distribution $\mathcal{N}(\sqrt{b_x^2+1/2},
(\omega_\mathrm{fluc}/2)^2)$. This agrees with the result of FOA
\cite{PhysRevB.65.205309}, since a high magnetic field suppresses spin
flips, leading to an Ising model and which features a Gaussian
distribution of polarization due to the central limit theorem.

\begin{figure}[tp]
\centering
\includegraphics[width=0.5\textwidth]{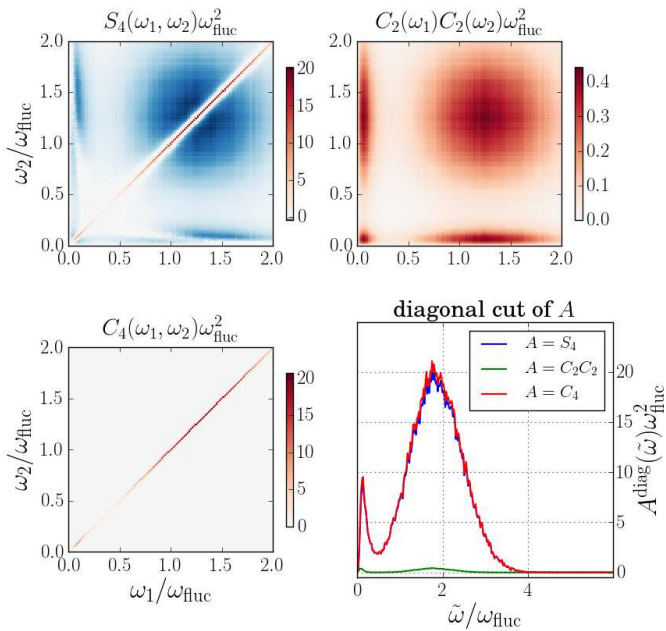} 
\caption{$S_4(\omega_1,\omega_2)\omega_\mathrm{fluc}^2$ as well as $C_4(\omega_1,\omega_2)\omega_\mathrm{fluc}^2$ and $C_2(\omega_1)C_2(\omega_2)\omega_\mathrm{fluc}^2$ for $b_\mathrm{ext}^x=1$ in the classical simulation with $N=100$ bath spins. In the lower right panel the diagonal cut  through all three spectra is shown.
}
\label{fig:bext}
\end{figure}

Combining the two contributions $C_4(\omega_1,\omega_2)$ and $C_2(\omega)$ leads to dominating
correlations on the diagonal $\omega_1=\omega_2$ as well as anticorrelations elsewhere in the $(\omega_1,\omega_2)$-plane
as  a consequence of the subtraction of both terms in Eq.\ \eqref{eq:S4_allg}.

To parametrize the diagonal cut we define $S_4^{\text{diag}}(\tilde
\omega) = S_4(\tilde{\omega}/\sqrt{2}, \tilde{\omega}/\sqrt{2})$ and
plot $S_4^{\text{diag}}(\tilde \omega)$ in the lower right panel of
Fig.\ \ref{fig:bext}. For small magnetic fields the spectrum changes
distinctively, as can be seen in Fig.~\ref{fig:bext}. Again we find a
Gaussian centred around $\sqrt{b^2_x+ 1/2}$ with a variance of
$(\omega_\mathrm{fluc}/2)^2$ but with reduced spectral weight.  For
$b_x=0$ the correlator $C_2$ features a strongly pronounced delta-peak
at $(\omega_1,\omega_2)=(0,0)$, as seen in Fig.\ \ref{fig:C2_comp},
with a maximum weight of one third of the total spectral weight in the
case of homogeneous coupling constants
\cite{PhysRevB.65.205309}. Increasing the strength of the external
magnetic field not only shifts the position of the Gaussian depending
on the external magnetic field but also transfers the weight of the
delta-peak to the Gaussian. For higher magnetic fields,
e.\,g. $b_x=5$, the contribution at $(0,0)$ has vanished, and only the
Gaussian remains. The same behavior also influences the $C_4$ part of
the spectrum, where we can observe a not yet disappeared delta-peak at
the origin of coordinates for $b_x=1$.

\begin{figure}[tp]
\centering
\includegraphics[width=0.45\textwidth]{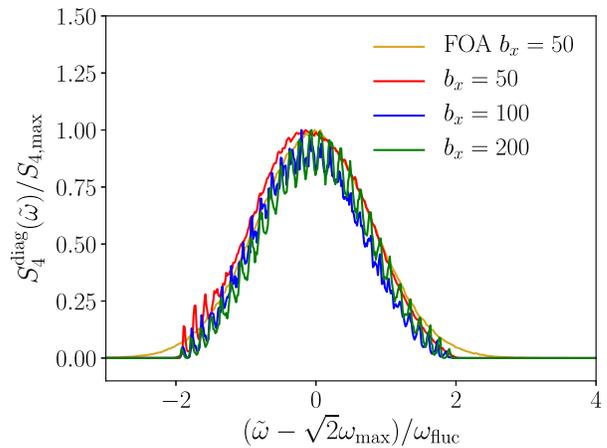} 
\caption{$S^{\text{diag}}_4(\tilde \omega)$ for high ($b_x=50, 100, 200$) transversal magnetic fields. Computed via the quantum mechanical scheme for  $N=3$ bath spins with a spin length of $I=9/2$. The spectra are shifted by $\sqrt{2}\omega_{\mathrm{max}}=\sqrt{2}\sqrt{b_x^2+1/2}$. The frozen Overhauser field approximation (FOA) is included for comparison.}
\label{fig:bhigh}
\end{figure}

After establishing the qualitative features of $C_4$ as well as the
product $C_2C_2$ for smaller and intermediate transversal field
strength $b_x$ by the classical simulation, we compare these results
with the quantum mechanical calculations for a very small bath but
with large nuclear in $I=9/2$ along the diagonal $\w_1=\w_2$.

$S_4^{\text{diag}}(\tilde \omega)$ for high magnetic fields is shown
in Fig.\ \ref{fig:bhigh}.  The definition can be used analogously for
the diagonal cuts through the $C_2(\omega_1)C_2(\omega_2)$ and
$C_4(\omega_1,\omega_2)$ spectra. While the quantum mechanical spectra
is centred around $\sqrt{b_x^2+ 1/2}$ at all magnetic fields and is
tracing the Gaussian envelope established in the classical simulation
for smaller fields, it develops a comb of peaks at high magnetic
fields.  At larger fields, spin-flip processes are suppressed, and the
dynamics becomes increasingly dominated by the Ising part of the CSM
in $x$-direction.  The peak location is governed by the hyperfine
interaction, with the distance decreasing with increasing bath sizes,
$\propto 1/\sqrt{N}$. The width of the peaks relates to the
variability of the $A_k$.  This phenomenon can not be observed with a
classical computation, where higher magnetic fields only shift the
spectrum which maintains its continuous shape.  With higher numbers of
bath spins and a distribution of $A_k$ with high variability, the
quantum mechanical spectrum would approach the results of the
classical simulation.

\subsubsection{Classical simulation as a limiting case of the quantum mechanical treatment of $S_4$}

\begin{figure}[tbp]
\begin{center}
\includegraphics[width=0.49\textwidth,clip]{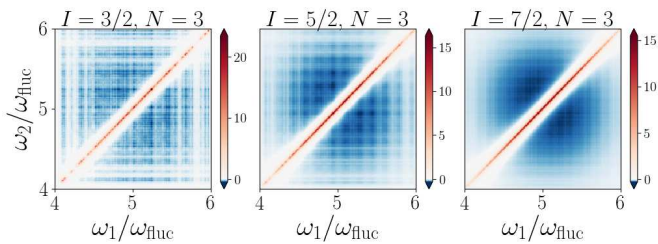}
\caption{$S_4(\omega_1, \omega_2)$ computed for a bath size of $N=3$ with spin lengths $I=3/2$, $I=5/2$ and $I=7/2$ and an $A_k$ configuration with $r_0=0.8$. The transversal magnetic field is set to $b_x=5$, and quadrupolar interaction is switched off.}
\label{fig:S4_Idiff}
\end{center}
\end{figure}

While the classical approach always yields a continuous frequency spectrum,
the variation of nuclear spin length as well as the bath size merits a more in-depth investigation for the quantum mechanical simulation.

\begin{figure}[tbp]
\begin{center}
\includegraphics[width=0.49\textwidth,clip]{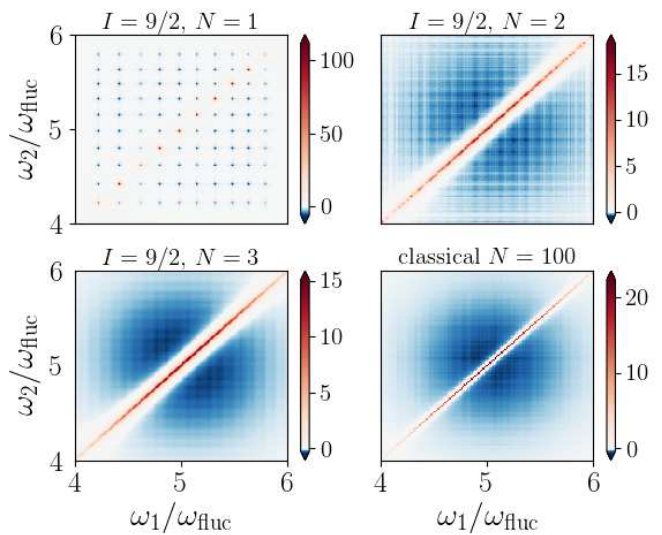}
\caption{$S_4(\omega_1, \omega_2)$ computed for $N=1, 2, 3$ bath spins with $I=9/2$, and for $N=100$ classical spins. The transversal magnetic field is $b_x=5$, and quadrupolar interaction is not included. }
\label{fig:S4_Ndiff}
\end{center}
\end{figure}

Figure \ref{fig:S4_Idiff} shows the quantum mechanical results for
$S_4(\omega_1, \omega_2)$ obtained by Eq.\ \eqref{eq:QM_S4} for
different spin lengths ($I=3/2,\, 5/2,\, 7/2$) and a fixed number of
bath spins ($N=3$) in a transversal field $b_x=5$ applying an average
over $N_a=32$ configurations of $\{ A_k\}$.  The spectrum becomes more
continuous with a growing spin length, due to the exponential increase
in the Hilbert space dimension and the larger number of non-degenerate
eigenenergies. As seen in the left panel of Fig.\ \ref{fig:S4_Idiff},
the non-zero contributions to $S_4$ are concentrated at a sparse
number of $(\omega_1, \omega_2)$ frequency pairs for $N=3$
$I=3/2$-spins, due to the limitations of the energy excitation
spectrum.  The delta-peaks in Eq.\ \eqref{eq:QM_S4} are broadened by a
factor $\gamma T^{\ast}=0.01$.  Correlations (red) are restricted to
the frequency subspace $\omega_1=\omega_2$, while the
anti-correlations (blue) can be found in an area centered around
$\omega_1=\omega_2\approx b_x$. Note the similarity between the
classical results (Fig.\ \ref{fig:S4_Ndiff}, lower right panel) and
the quantum mechanical solution for $N=3$ and $I=7/2$ (Fig.\
\ref{fig:S4_Idiff}, right panel), solidifying the conjecture that the
quantum mechanical spectra approaches the results of the classical
simulation in the limit of $I\to\infty$.

The fourth-order cumulant spectra $S_4(\omega_1, \omega_2)$ are
presented for different $N$ and a fixed spin length $I=9/2$ at $b_x=5$
in Fig.\ \ref{fig:S4_Ndiff}.  For $N=1$, the position of the non-zero
contributions are clearly governed by the Zeeman splitting of the
nuclear spins coupled to the central spin via a single hyperfine
coupling constant $A=\w_{\rm fluc}$.  This results in $(2I+1)^2$
equidistant peaks on a grid around the point given by $(\omega_L,
\omega_L)$, that are positive at the $\omega_1=\omega_2$ diagonal and
negative everywhere else.  For larger bath sizes the spectrum becomes
more continuous. At $N=3$ bath spins of length $I=9/2$ the $S_4$
spectrum, as displayed in the lower left panel of Fig.\
\ref{fig:S4_Ndiff}, is already qualitatively very similar to the
classical result depicted on the lower right panel of Fig.\
\ref{fig:S4_Ndiff}.

The simulations show that the classical calculations are valid limits
of the quantum mechanical calculations for $I\rightarrow \infty$ and
$N\rightarrow \infty$.  Furthermore, we established that fourth order
cumulant does not vanish implying that the central spin does not
behave as a classical random variable whose noise spectrum is purely
of Gaussian type. The physics is driven by the coherent precession
around the external constant magnetic field in combination with a
slowly varying nuclear spin dynamics.  The FOA reveals the restriction
of $C_4$ to the frequency diagonal which is shared by both approaches
that explicitly include the nuclear spin dynamics.

\subsection{Influence of quadrupolar interaction on $S_4(\omega_1, \omega_2)$}
\label{sec:Quad_Bench}

Within the CSM, the positive correlations are restricted to the
diagonal $\w_1=\w_2$ related to the spectral confinement of
$C_4(\w_1,\w_2)$ leading to anti-correlation everywhere else in the
frequency plane. In this section, we add the nuclear-electric
quadrupolar interaction $H_Q$ to the CSM and investigate its influence
onto $S_4$.

\begin{figure}[tbp]
\begin{center}
\includegraphics[width=0.49\textwidth,clip]{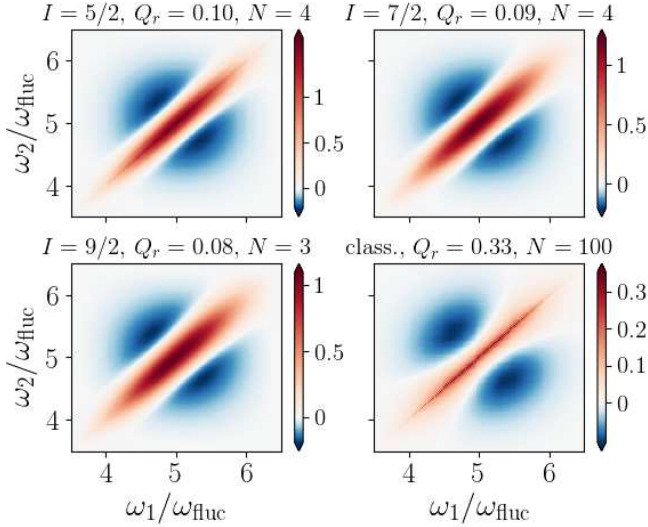}
\caption{$S_4/\omega_1, \omega_2)$ with a magnetic field $b_\mathrm{ext}^x=5$, with different spin lengths $I$, bath sizes $N$ and interaction strengths $Q_r$. The parameters are chosen for similar behavior in the second order spin correlation, see Fig.\ \ref{fig:C2_Idiff}.}
\label{fig:S4_Idiff_quad}
\end{center}
\end{figure}

\subsubsection{Fourth-order spin noise at intermediate and large magnet fields}

Here, we focus on intermediate and large magnet fields since in this
regime the spin-noise power spectrum $C_2(\w)$ remains unaltered in
the presence of quadrupolar interaction. In leading order $C_2(\w)$ is
described by a Gaussian \cite{PhysRevB.65.205309,Glazov2012} centered
around $\omega_1$ -- see also Sec.\ \ref{sec:C2_noise_benchmark}.

Figure \ref{fig:S4_Idiff_quad} shows $S_4(\omega_1, \omega_2)$
computed quantum mechanically for bath spin lengths $I=3/2,\, 7/2,\,
9/2$ as well as the results of the classical approach. The strength of
the quadrupolar coupling is chosen such that $C_2(t)$ agrees for
$b_{\text{ext}}=0$ independent of the spin length -- see the
discussion in Sec.\ \ref{sec:C2_noise_benchmark}. While the
fourth-order contribution to $S_4$ is restricted to the diagonal,
$\omega_1=\omega_2$ without quadrupolar interaction, the introduction
of quadrupolar couplings causes a broadening of the heretofore sharp
peak. But while the quantum mechanical cumulant spectra look very
similar, the classically computed $S_4$ exhibits a much smaller
broadening of the positive contribution around $\omega_1=\omega_2$,
and a qualitatively different peak shape as can be seen in the lower
right panel of Fig.\ \ref{fig:S4_Idiff_quad}.  Since the quadrupolar
interaction does not affect the shape of $C_2(\omega)$ for transversal
magnetic fields $b_{x} > 1$ in both approaches, the mismatch between
quantum mechanical and classical fourth-order cumulant is related to
$C_4$.

\begin{figure}[tbp]
\begin{center}
\includegraphics[width=0.49\textwidth,clip]{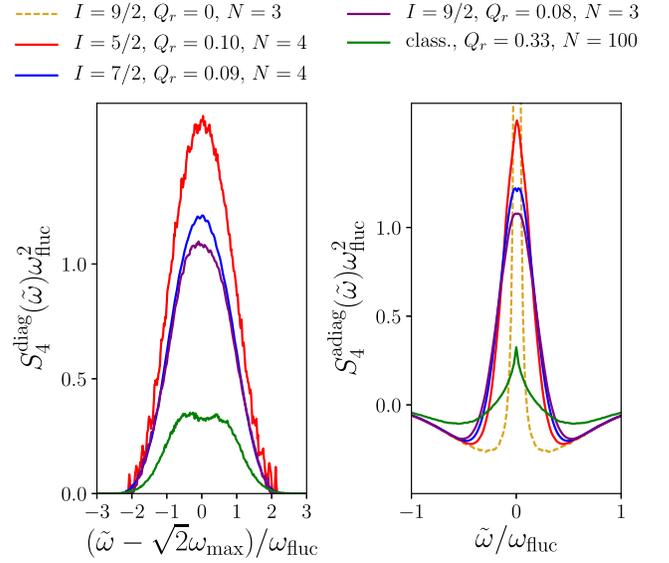}
\caption{The same data plotted in Fig.\ \ref{fig:S4_Idiff_quad}, cut in the diagonal $\omega_1=\omega_2$, $S_4^{\text{diag}}(\tilde \omega)$,  as well as in the anti-diagonal cut $\omega_1+\omega_2=2\omega_{\mathrm{max}}$, $S_4^{\text{adiag}}(\tilde \omega)$. $S_4^{\text{diag}}(\tilde \omega)$ without quadrupolar coupling was added in the right panel for comparison.}
\label{fig:S4_Idiff_cut}
\end{center}
\end{figure}

To further investigate the broadening of the correlation caused by
quadrupolar coupling, especially how this broadening behaves dependent on the $Q_r$, we analyze the broadening of $C_4$ perpendicular
to the frequency diagonal as function of $Q_r$. For that purpose, we
parametrize the anti-diagonal cut in the vicinity of its global
maximum, $S_4(\omega_{\text{max}}, \omega_{\text{max}})$, with
$\omega_{\text{max}}/\omega_{\text{fluc}}=\sqrt{b^2_x+1/2}$ by
$\omega_1+\omega_2=2\omega_{\text{max}}$.  We define the corresponding
anti-diagonal cut as
\begin{align}
S_4^{\text{adiag}}(\tilde \omega)= S_4\left(\omega_{\text{max}}+\frac{\tilde \omega}{\sqrt{2}}, \omega_{\text{max}}-\frac{\tilde \omega}{\sqrt{2}}\right)
\end{align}
so that the global maximum is located at the relative frequency $\tilde \omega = 0$.

The diagonal and anti-diagonal cuts of the data presented in Fig.\
\ref{fig:S4_Idiff_quad} are plotted in Fig.\ \ref{fig:S4_Idiff_cut}.
The same parameters that produced congruent results for conventional
spin noise spectrum $C_2(\w)$ as shown in Fig.\ \ref{fig:C2_comp}, now
lead to markedly different behaviour.  The left panel shows the
diagonal cuts computed with the quantum mechanical method for
different nuclear spin length I and bath size $N$ and is augmented by
the results of the classical approach for $N=100$ nuclear spins.  The
diagonal cuts exhibit roughly the same Gaussian behaviour independent
of $Q_r$, but its amplitude decreases by about a factor five.  This is a direct result of the broadening observed in
Fig.\ \ref{fig:S4_Idiff_quad}, as the total spectral weight of $C_4$
as well as $S_4$ remains conserved.  The drop in amplitude is not
uniform, but is more pronounced in $S_4$ computed via the classical
approach, suggesting that the quadrupolar coupling has a stronger
effect there.  In the quantum mechanically computed $S_4$ the
amplitude decreases with larger $I$.

On the right panel of Fig.\ \ref{fig:S4_Idiff_cut} the anti-diagonal
cuts are shown for the same parameters as in the left panel. Added for
comparison is $S_4^{\text{adiag}}(\tilde \omega)$ for $Q_r=0$
obtained with the same broadening parameter $\gamma$.
$S_4^{\text{adiag}}(\tilde \omega)$ reveals a fundamentally different
curve shape depending on the computational approach.  While the
classical curve exhibits a cusp which could be fitted by a power law,
the quantum mechanical approach yields a Gaussian shape.

\begin{figure}[tbp]
\begin{center}
\includegraphics[width=0.49\textwidth,clip]{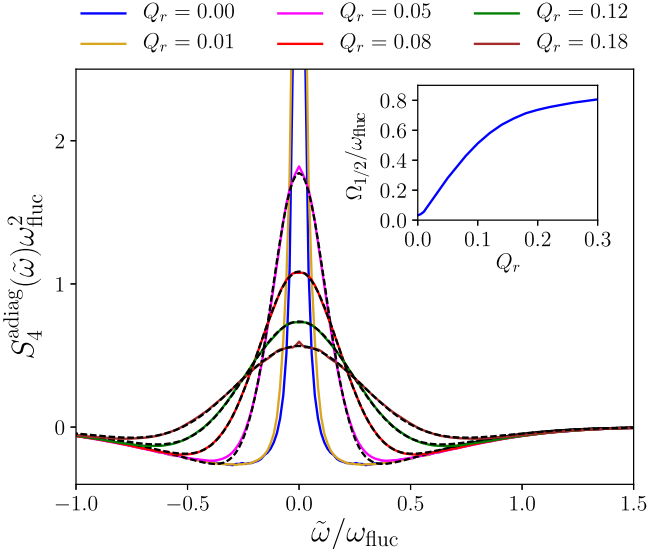}
\caption{$S_4^{\text{adiag}}(\tilde \omega)$ quantum mechanically calculated with $N=3$, $b_x=5$ and $I=9/2$ for different $Q_r$. The inset plot shows the full width half maximum $\Omega_{1/2}$ of $C_4$ in relation to the quadrupolar coupling strength $Q_r$.}
\label{fig:S4_Qdiff_cut}
\end{center}
\end{figure}

The scaling behavior which allows us to match classical and quantum
mechanical results for $C_2(\w)$ by adjusting $Q_r$,
cf.\,Sec. \ref{sec:C2_noise_benchmark}, therefore only holds for the
second order spin noise and not the fourth order spin-noise
bispectrum. It stands to reason that the quantum mechanical method
includes features that have been neglected in the classical approach,
such as the non-commutability of the bath spin components.

In order to connect the relative quadrupolar coupling strength $Q_r$
with the broadening of the anti-diagonal, we plotted
$S_4^{\text{adiag}}(\tilde \omega)$ for different $Q_r$ and fixed
$N=3$ and $I=9/2$ in Fig.\ \ref{fig:S4_Qdiff_cut}.  The contribution
$C_2(\omega_1)C_2(\omega_2)$, can be represented by a Gaussian with
the variance $\sigma^2=(\omega_\mathrm{fluc}/2)^2$ independent of
$Q_r$ compatible with the FOA \cite{PhysRevB.65.205309}.  The
fourth-order contribution $C_4$ in contrast changes markedly with the
quadrupolar interaction strength.  Fitting only $C_4$ with a Gaussian
leads to the relation between the full-width half maximum
$\Omega_{1/2}$ and the quadrupolar coupling strength $Q_r$ shown in
the inset of Fig.\ \ref{fig:S4_Qdiff_cut}. For small $Q_r$, the
dependence is roughly linear, before the increase flattens at
$Q_r>0.1$. For $Q_r \rightarrow 0$, the Gaussian curve becomes a sharp
peak $\Omega_{1/2}\rightarrow 0$ limited here due to the Lorentz
broadening simulating a finite measuring time $T_m$.

\begin{figure}[t]
\begin{center}
\includegraphics[width=0.49\textwidth,clip]{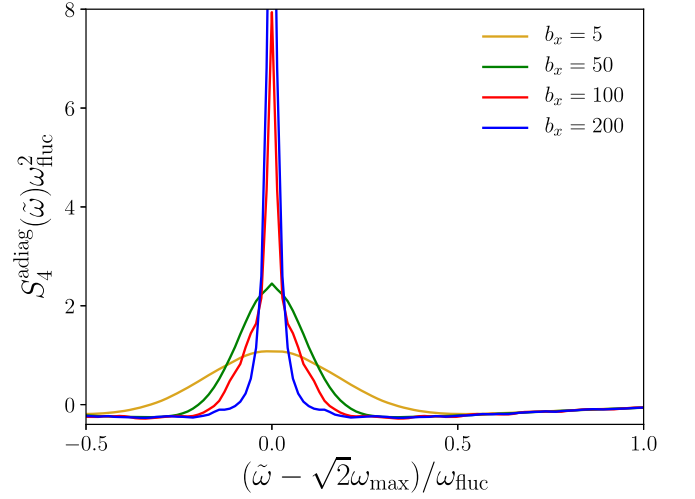}
\caption{$S_4^{\text{adiag}}(\tilde \omega)$ quantum mechanically calculated with $N=3$, $Q_r=0.08$ and $I=9/2$ for different transversally applied external magnetic fields $b_x$. }
\label{fig:S4_cut_Q_Bdiff}
\end{center}
\end{figure}

Figure\ \ref{fig:S4_cut_Q_Bdiff} depicts the anti-diagonal cut
$S_4^{\text{adiag}}(\tilde \omega)$ for different magnetic fields
$b_x$ and fixed spin bath size and spin length. The quadrupolar
coupling induced broadening decreases with an increasing magnetic
field strength: the dynamics of the system is dominated by the Zeeman
energy, and $H_Q$ becomes an increasingly weaker perturbation. This
agrees well with the observation of the fourth-order spin correlation
function in the time domain \cite{PhysRevB.96.045441}, where the high
magnetic fields shift the decay time from $\mathcal{O}(\mathrm{ns})$
to an exponential decay with a magnetic field dependent decay time
$T_2\propto \mathcal{O}(\mu\mathrm{s})$
\cite{Press2010,Bechtold2016,FinleyNature}.

\begin{figure}[tbp]
\begin{center}
\includegraphics[width=0.49\textwidth,clip]{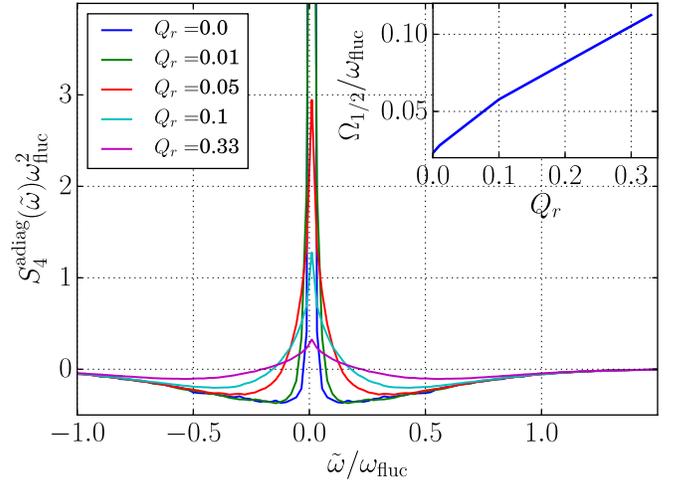}
\caption{Cuts for the classically calculated $S_4^{\text{adiag}}(\tilde \omega)$. The inset plot shows the full width at half maximum $\Omega_{1/2}$ of the $C_4$ part dependent on the quadrupolar coupling strength $Q_r$.  }
\label{fig:S4_Qdiff_cut_class}
\end{center}
\end{figure}

We performed the same type of simulations as in Fig.\
\ref{fig:S4_Qdiff_cut} using the classical approach.  Figure
\ref{fig:S4_Qdiff_cut_class} shows $S_4^{\text{adiag}}(\tilde \omega)$
for different $Q_r$.  Since $C_2(\omega)$ remains invariant under the
change of $Q_r$, the change in the spectrum is directly linked to the
change of $C_4(\w_1,\w_2)$. As in the quantum mechanical simulations,
the quadrupolar interaction lifts the spectral constrain to
$\w_1=\w_2$ in $C_4(\w_1,\w_2)$. The overall sum-rule for
$C_4(\w_1,\w_2)$ implies a decrease of the peak at $\tilde \w=0$ and
an increasing distribution of spectral weight into the $(\w_1,\w_2)$
plane. Since the shape of the classical $S_4^{\text{adiag}}(\tilde
\omega)$ is non-Gaussian, we extracted the full-width half maximum
$\Omega_{1/2}$ of $C_4$ as function of $Q_r$ and plotted the result as
inset in Fig.\ \ref{fig:S4_Qdiff_cut_class}. In full agreement with
the quantum mechanical approach we find a linear dependency of
$\Omega_{1/2}$ on $Q_r$. The finite offset at $Q_r=0$ is related to
the finite size effect of the Fourier transformation for
$T_m<\infty$. The absolute value of $\Omega_{1/2}$, however, differs
between the quantum mechanical and the classical simulations which we
attribute to the bath size difference.

\subsubsection{Fourth-order spin noise in the crossover regime}

\begin{figure}[tbp]
\begin{center}
\includegraphics[width=0.23\textwidth,clip]{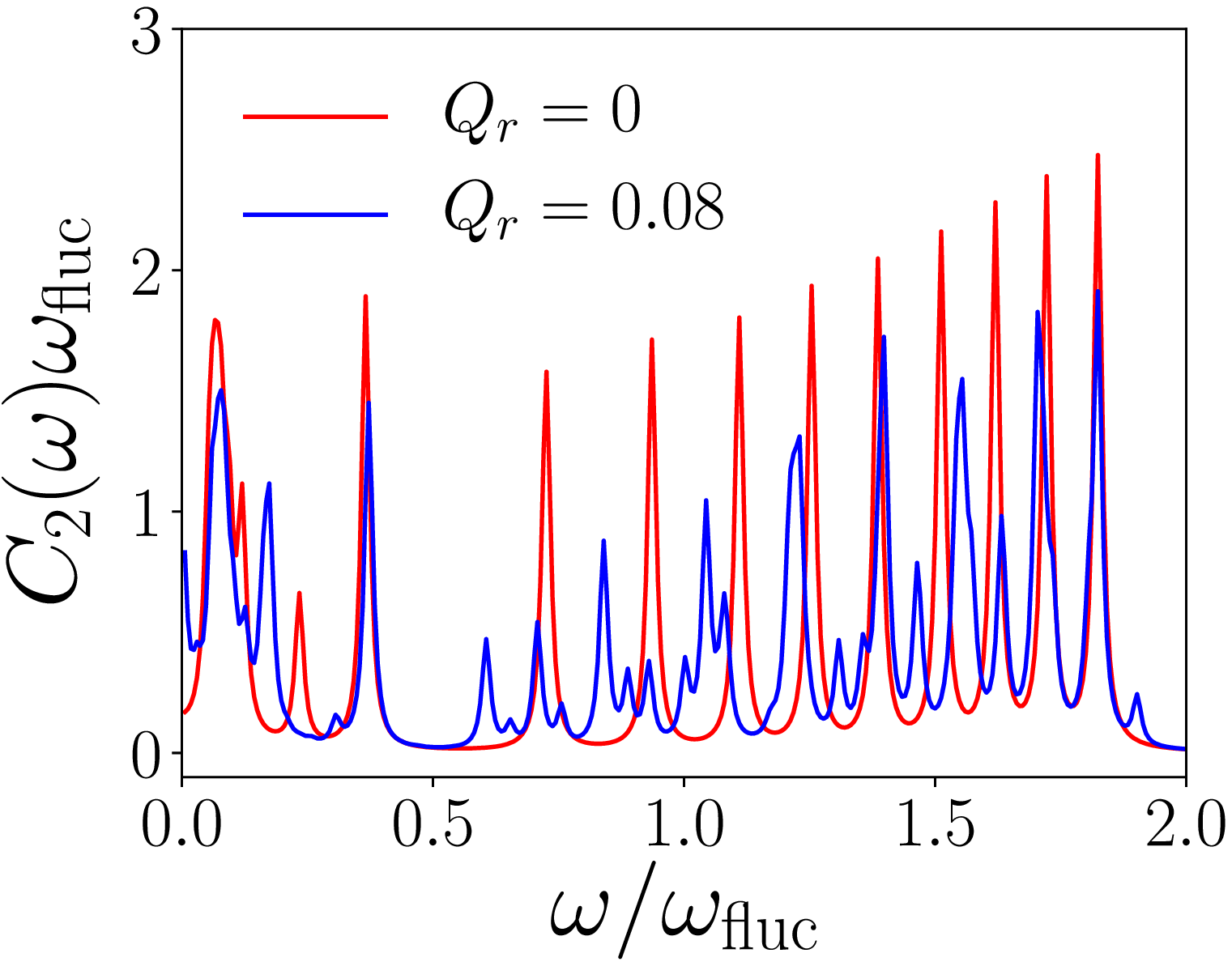}
\includegraphics[width=0.23\textwidth,clip]{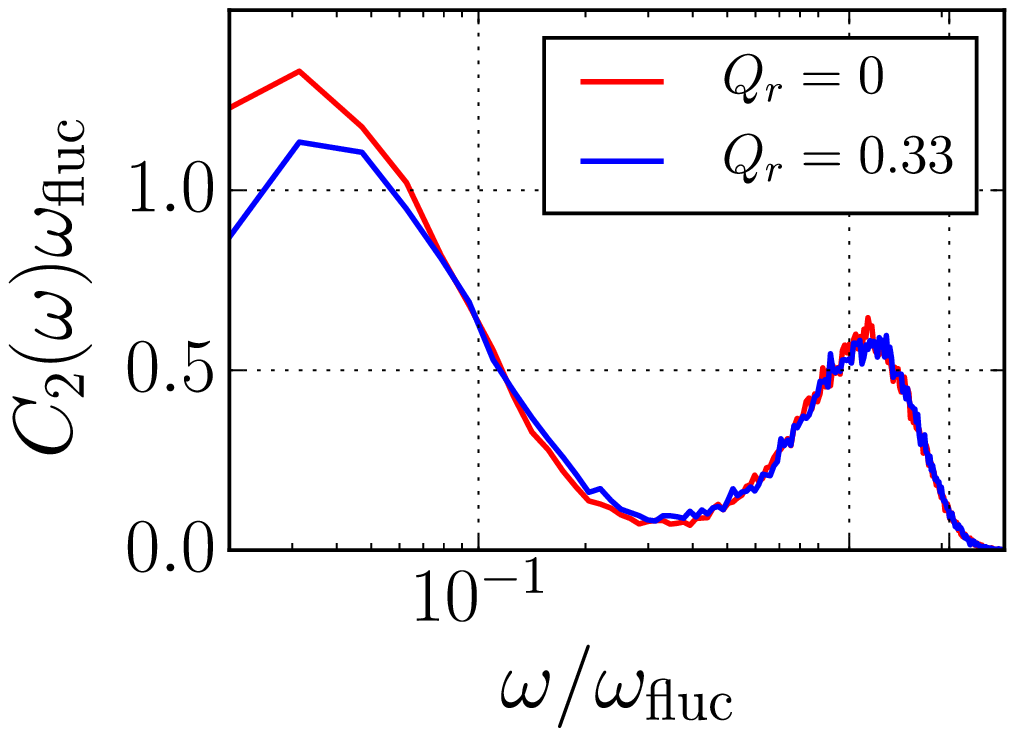}
\caption{Comparison of $C_2(\w)$ with and without $H_Q$. The left panel shows
the results of quantum mechanical calculation, with $N=1$, $I=9/2$ and $Q_r=0.08$,
the right panel the classically computed spectrum using $N=100$, $Q_r=0.33$.}
\label{fig:c2-hagele}
\end{center}
\end{figure}

Now we turn to the crossover regime where the Zeeman energy is of the
order of $\w_{\rm fluc}$, i.\ e.\ $b_x\approx 1$. The electron spin
dynamics is governed by the external magnetic field and the
fluctuating Overhauser field which have equal strength. Furthermore,
the nuclear Zeeman energy is weak such that the nuclear spin dynamics
is dominated by the nuclear-electric quadrupolar interaction in
combination with the weak Knight field generated by the electron
spin. We are interested in comparing two extreme limits: the dynamics
of the smallest system one can imagine, including only a single
nuclear spin, and the limit of large number of spins. While $N=1$
requires a purely quantum mechanical calculations, we mimic the large
N limit with a classical simulation of $N=100$ bath spins.

In this regime $H_Q$ does not only influence $S_4$ but also modifies
$C_2(\w)$.  The change of $C_2(\w)$ induced by the quadrupolar
interaction is depicted in Fig.\ \ref{fig:c2-hagele} for bath sizes
$N=1$ (left panel) and $N=100$ (right panel).

\begin{figure}[tbp]
\begin{center}
\includegraphics[width=0.22\textwidth,clip]{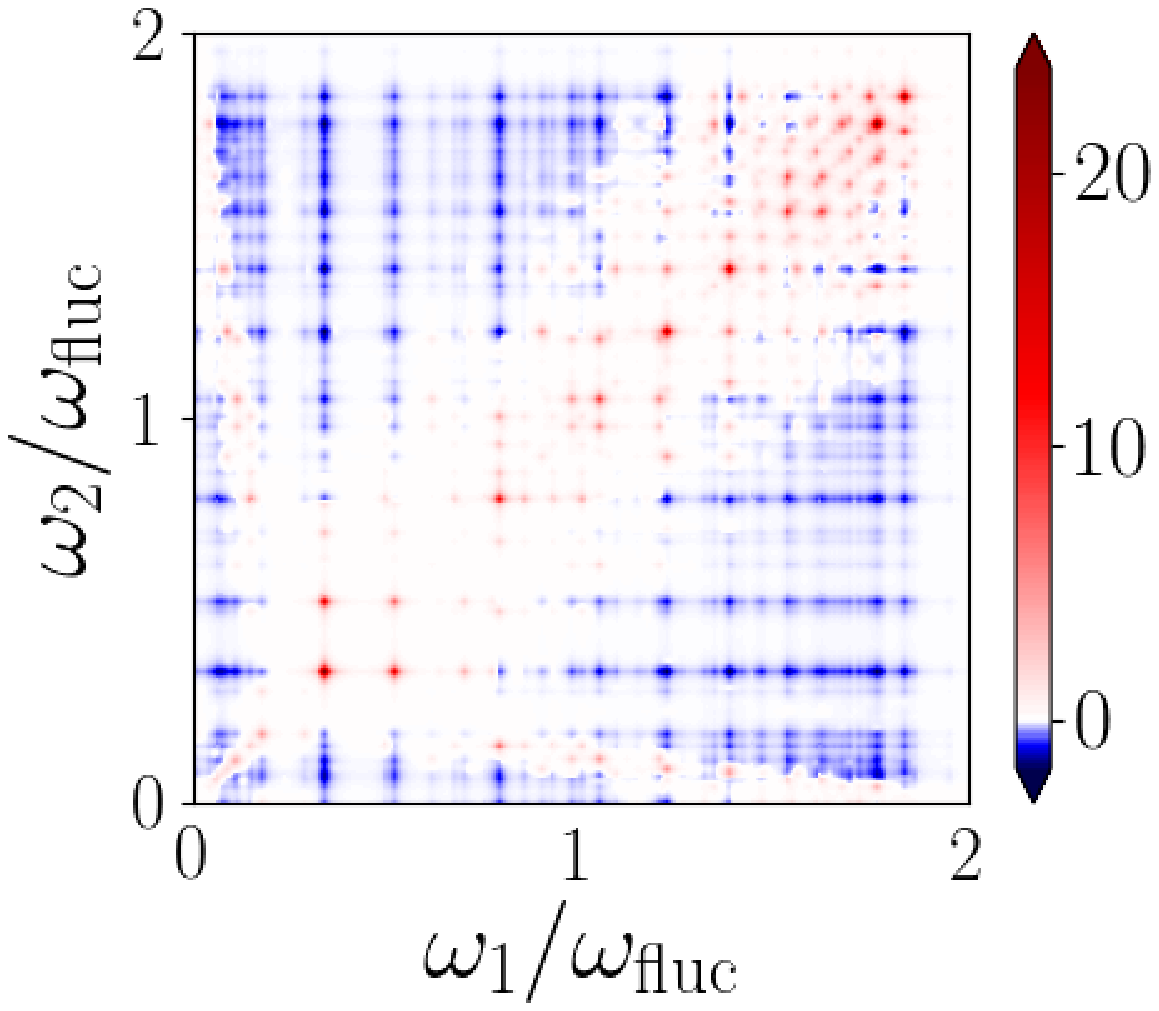}
\includegraphics[width=0.24\textwidth,clip]{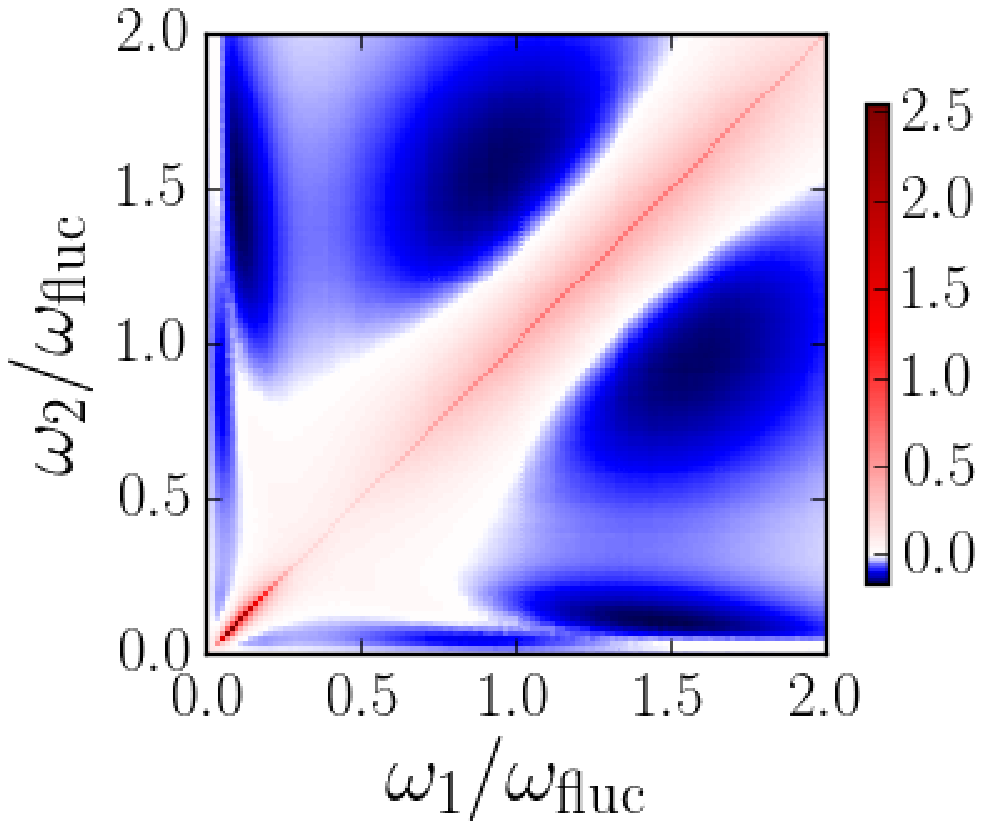}
\caption{$S_4$ for a smaller magnetic field of $b_x=1$. Left is the quantum mechanical calculation, with $N=1$, $I=9/2$ and $Q_r=0.08$. On the right is the classically computed spectrum with $N=100$, $Q_r=0.33$.}
\label{fig:hagele}
\end{center}
\end{figure}

We use the corresponding spin noise $C_2$ to calculate the fourth
order cumulant.  In Fig.\ \ref{fig:hagele} the classical and the
quantum mechanical results of $S_4$ are presented for $b_x=1$. Note
that the quantum mechanical $S_4$ on the left computed in the limit of
weak measurement for $N=1$, is near identical to the $S_4$ presented
for strong and continuous measurement in Ref.\
\cite{PhysRevB.98.205143}.  For small $\w_1$ or $\w_2$ we also found
alternating signs of correlations in the $(\w_1,\w_2)$-plane.  Fixing
$\w_1/\w_{\rm fluc}=1$ and increasing $\w_2$ reveals first weak
anti-correlation (encoded in blue), then correlations (encoded in red)
before switching back to anti-correlations. Also, the strong
correlations are not confined to the diagonal as depicted in Fig.\
\ref{fig:S4_Ndiff} but are significantly spread due to the presence of
the quadrupolar couplings.

The classically obtained $S_4$ on the right shows the effects of
quadrupolar coupling in small magnetic fields for a far bigger bath of
$N=100$, which results in a continuous spectrum with similar
features. These are the anti-correlation contributions at the axis
with a dip in anti-correlation along $\omega_{1/2}/\w_{\rm fluc}=1$,
as well as the broadening of the correlation on the diagonal.

\subsubsection{Discussion}

While the second-order spin correlation function $C_2(t)$ decays fast
on the time scale $T^*$ in finite magnetic field that
long-time effects of nuclear quadrupolar coupling cannot be observed
in the electron spin dynamics, they modify the frequency
characteristics of the fourth-order spin correlation function
significantly. The positive correlations in the spin noise power
bispectrum that are pinned to the frequency diagonal in the CSM are
broadened and acquire a finite width proportional to the nuclear
coupling strength at a large magnetic field.

With and without quadrupolar interaction the classical and the quantum
mechanical method yield congruent results for the second-order
correlation. The same is true for the fourth-order correlation without
quadrupolar interaction. If quadrupolar interaction is introduced,
both the classical and quantum mechanical method show qualitatively
similar behavior - a broadening of the correlation peak along the
$\omega_1+\omega_2=\mathrm{const}$ cut. But, as can be seen in Fig.\
\ref{fig:S4_Idiff_cut}, the spectra of different methods exhibit a
quantitatively different curve progression, and to not follow the same
scaling behavior presented in Sec.\ \ref{sec:C2_noise_benchmark}. This
shows that the fourth-order correlation yields uniquely quantum
mechanical information that appears with the introduction of
quadrupolar interaction into the system, as has been previously shown
in \cite{PhysRevB.96.045441}.

It is straight forward to extend the investigation to an arbitrary
angle between the $z$-axis and the applied magnetic field. This is a
well studied problem in the context of the standard SNS and it turns
out that a tilted magnetic field does not provide additional new
information. Therefore, we do not include these results in this
paper. For the limiting case of a single bath spin, i.\ e.\ $N=1$, we
refer to Fig.\ 6 in Ref.\ \cite{PhysRevB.98.205143} which extrapolated
to continuous spectra as obtained with our classical simulation.

\section{Conclusion}
\label{sec:Conclusion}

We presented a combination of a quantum mechanical and a classical
simulation to the fourth order noise correlation function, to
calculate the spin-noise power bispectrum in a quantum dot in the
presence of the nuclear-electric quadrupolar interaction in the limit
of a very small and a very large nuclear spin bath. Our approach is
valid in the limit of a nearly perturbation free off-resonance
detection of the spin polarisation in the quantum dot ensemble using
the Faraday rotation of a weak linear polarized optical probe signal.

The second-order spin correlation function $C_2(t)$ is used as a gauge
to connect the nuclear spin length $I$ and the effective quadrupolar
interaction strength $Q_r$ to the number of nuclear spins of the spin
bath in all calculations. To account for the effect of quadrupolar
interaction in a classical spin dynamics, we derived a modification of
the effective Knight field in classical equations of motions.

The quantum mechanical and the classical spin-noise bispectrum agree
well for the CSM.  The quantum mechanical bispectrum converges to the
result of the classical simulation for large nuclear spin bath and
large nuclear spin length. In both cases the quantum mechanical
eigenvalue spectrum approaches a continuum
distribution. Interestingly, already relatively small spin baths
provide a good representation of a larger bath bispectrum.

 The fourth order cumulant $S_4$ is made up of two basic building
blocks: $C_4(\omega_1,\omega_2)$ and $C_2(\omega_1)C_2(\omega_2)$. The
decomposition of those parts show that the product of the second-order
spin noise gives a 2D Gaussian which is solely responsible for
anti-correlation in the spectrum while $C_4$ is non-zero only on the
diagonal in the CSM.

Adding the quadrupolar interaction term $H_Q$ to the CSM is causing a
broadening of $C_4$ across the diagonal. The width of this broadening
is directly proportional to the quadrupolar coupling strength at small
couplings and a finite magnetic field. The width could be used as a
direct experimental probe of the average quadrupolar interaction
strength in a sample.  The near perfect agreement observed in
$C_2(\omega)$ between the classical and the quantum mechanical
simulations is slightly modified in the bispectrum. The qualitative
agreement between the bispectra of both methods with comparable
parameters is remarkable concerning the location of the correlation as
well as the anti-correlations.  The broadening of the quantum
mechanical spectra $C_4$ along the diagonal, however, is more
pronounced than in its classical counterpart, while the decrease of
the amplitude due to quadrupolar interaction is stronger for the
results of the classical method.  The difference in the shape between
the results of both methods becomes visible in the cut through the
diagonal.

We have proven that the simple linear response theory to higher
correlation functions \cite{1367-2630-15-11-113038,LiSinitsyn2016}
produces congruous results to those obtained with an elaborate weak
measurement theory presented in Ref.\ \cite{PhysRevB.98.205143}. This
shows that the assumption of a non-perturbative measurement yields
identical results that the weak measurement theory in the weak
coupling limit \cite{PhysRevB.98.205143}.

\begin{acknowledgements}
We acknowledge the financial support by the
Deutsche Forschungsgemeinschaft and the Russian Foundation of Basic
Research through the transregio TRR 160 project A7, and we thank Daniel H\"agele and Manfred Bayer for the fruitful discussions.
\end{acknowledgements}

\appendix


%

\end{document}